\documentclass[10pt,conference]{IEEEtran}

\IEEEoverridecommandlockouts

\usepackage{cite}
\usepackage{amsmath,amssymb,amsfonts}
\usepackage{algorithm}
\usepackage{algorithmic}
\usepackage{graphicx}
\usepackage{textcomp}
\usepackage{xcolor}
\usepackage{optidef} 
\usepackage{verbatim}
\usepackage{todonotes}
\usepackage{caption}
\usepackage{subcaption}

\begin{document}
\title{\Large{Joint User Pairing and Beamforming Design of Multi-STAR-RISs-Aided\\NOMA in the Indoor Environment via Multi-Agent Reinforcement Learning}}
\author{\IEEEauthorblockN{Yu~Min~Park$^1$,~Yan~Kyaw~Tun$^2$,~and~Choong~Seon~Hong$^1$}
\IEEEauthorblockA{$^1$Department of Computer Science and Engineering, Kyung Hee University, Yongin, 17104, Republic of Korea\\
$^2$Department of Electronic Systems, Aalborg University, . C. Meyers Vænge 15, 2450 København}
Email: \{yumin0906, cshong\}@khu.ac.kr, ykt@es.aau.dk.}

\maketitle
\begin{abstract}
The development of sixth-generation (6G)/Beyond Fifth-Generation (B5G) wireless networks, which have requirements that go beyond current 5G networks, is gaining interest from academia and industry. However, to increase 6G/B5G network quality, conventional cellular networks that rely on terrestrial base stations are constrained geographically and economically. Meanwhile, Non-Orthogonal Multiple Access (NOMA) allows multiple users to share the same resources, which improves the spectral efficiency of the system and has the advantage of supporting a larger number of users. Additionally, by intelligently manipulating the phase and amplitude of both the reflected and transmitted signals, Simultaneously Transmitting and Reflecting RISs (STAR-RISs) can achieve improved coverage, increased spectral efficiency, and enhanced communication reliability. However, STAR-RISs must simultaneously optimize the amplitude and phase shift corresponding to reflection and transmission, which makes the existing terrestrial networks more complicated and is considered a major challenging issue. Motivated by the above, we study the joint user pairing for NOMA and beamforming design of Multi-STAR-RISs in an indoor environment. Then, we formulate the optimization problem with the objective of maximizing the total throughput of mobile users (MUs) by jointly optimizing the decoding order, user pairing, active beamforming, and passive beamforming. However, the formulated problem is a mixed-integer non-linear programming (MINLP). To address this challenge, we first introduce the decoding order for NOMA networks. Next, we decompose the original problem into two subproblems, namely: 1) MU pairing and 2) Beamforming optimization under the optimal decoding order. For the first subproblem, we employ correlation-based K-means clustering to solve the user pairing problem. Then, to jointly deal with beamforming vector optimizations, we propose Multi-Agent Proximal Policy Optimization (MAPPO), which can make quick decisions in the given environment owing to its low complexity. Finally, simulation results prove that our proposed MAPPO algorithm is superior to Proximal Policy Optimization (PPO) and Advanced Actor-Critic (A2C) by a maximum of 1\% and 6\%, respectively.\\

\end{abstract}
\begin{IEEEkeywords}
STAR-RIS, NOMA network, indoor environment, reinforcement learning, multi-agent proximal policy optimization.
\end{IEEEkeywords}

\section{Introduction}

There is a growing interest among both academic and industrial circles regarding the advancement of sixth-generation (6G)/Beyond Fifth-Generation (B5G) wireless networks. The requirements are to address the more stringent demands that surpass those of the existing 5G networks. These requirements include achieving ultra high data rates and energy efficiency, ensuring global coverage and connectivity, and attaining extremely high reliability and low latency \cite{chen2022reconfigurable, tun2019wireless}. However, Existing cellular networks that rely on terrestrial base stations have economic and geographic limitations for improving network quality. Meanwhile, Reconfigurable Intelligent Surfaces (RISs) are new communication equipment for future next-generation wireless communication network performance improvement \cite{STAR_RIS_lit_1}. RIS is a plane reflector composed of multiple low-cost reconfigurable passive communication elements. The corresponding element may reconstruct the radio signal propagation by manually adjusting the amplitude and phase appropriately. Therefore, RIS may be deployed in a wireless network concentration area to improve communication quality with economical and low energy consumption. In \cite{park2022trajectory}, the authors minimized the latency by improving the communication throughput of ground users through RIS located in buildings in full-duplex communication environments. In addition, the work in \cite{park2022joint} provided a study that provides wireless communication for high-speed trains by further maximizing line-of-sight (LoS) by mounting RIS on UAVs. However, the disadvantage of existing RIS is that it only has a reflection function, so the transmitter and receiver must be on the same side. This topological constraint limits the flexibility of employing existing RISs.

To overcome this, unlike the RIS described above, Simultaneously Transmitting and Reflecting (STAR-RIS) can provide communication services to both parties by enabling simultaneous transmission and reflection of incident signals. STAR-RIS is largely classified into three types depending on how transmission and reflection signals are controlled \cite{liu2021star}. The types of STAR-RIS are classified into Energy Splitting (ES), which controls the energy for transmission and reflection signals, Mode Switching (MS), which converts the mode of each element to determine the signal method, and Time Switching (TS), which changes the signal method over time. Among these types, ES-type STAR-RIS has high flexibility but has the disadvantage of optimizing energy variables along with the phase variables of each signal. By intelligently manipulating the phase and amplitude of both the reflected and transmitted signals, STAR-RISs can achieve improved coverage, increased spectral efficiency, and enhanced communication reliability \cite{STAR_RIS_lit_4}. Meanwhile, in \cite{wang2023joint}, a study was conducted on how to optimize a number of STAR-RISs. However, optimization for the Non-Orthogonal Multiple Access (NOMA) networks was left as an assumption without considering it.

NOMA is a multi-access technology that allows multiple users to access the same frequency band simultaneously without the need for orthogonal resource allocation, such as the existing Orthogonal Frequency Division Multiple Access (OFDMA) or Code Division Multiple Access (CDMA) systems \cite{islam2018resource}. NOMA can separate signals that overlap each other in the same time and frequency domain from the receiver using continuous interference cancellation (SIC) or other signal processing techniques. This allows multiple users to share the same resources, which improves the spectral efficiency of the system and has the advantage of supporting a larger number of users. However, there are currently no studies that have carefully addressed the Multi-STAR-RISs-aided NOMA networks. In this paper, we try to fill this gap and our major contributions may be summed up as follows:
\begin{itemize}
    \item We propose a novel network architecture for indoor environment wireless communication where the access point provide services to multiple mobile users with the aid of multiple STAR-RIS (Multi-STAR-RISs) in NOMA networks.
    \item Drawing upon the suggested system architecture, we formulate an optimization problem to address the user pairing, AP active beamforming, as well as passive beamforming included amplitude and phase shift of STAR-RISs.
    \item To tackle the aforementioned problem, we decompose the main problem into two sub-problems by using the block coordinate descent (BCD) method, and then solve each sub-problem, iteratively.
    \item Additionally, we propose correlation-based clustering to address the user pairing problem. In addition, the beamforming opimization problem was solved by applying multi-agent reinforcement learning (MARL) to optimize beamforming vectors.
    \item Finally, we perform in-depth simulations to show that our suggested approach performs better than the baseline algorithms. The simulation results show that the proposed MAPPO performance is better than comparison algorithms to a maximum of 6\%.
\end{itemize}

The subsections of the paper are organized in the following manner. The system model and problem formulation are outlined in Section \ref{SF}. Subsequently, Section \ref{sol_algo} provides a detailed description of the method that has been proposed. The details of the implementation and simulation outcomes are expounded upon in Section \ref{simul}, while the conclusion of the paper comes in Section \ref{conclusion}.

\section{System Model \& Problem Formulation}
\label{SF}
\subsection{System Model Overview}
\label{system_model}
\begin{figure}[h!]
    \centering
    \includegraphics[width=7.5cm]{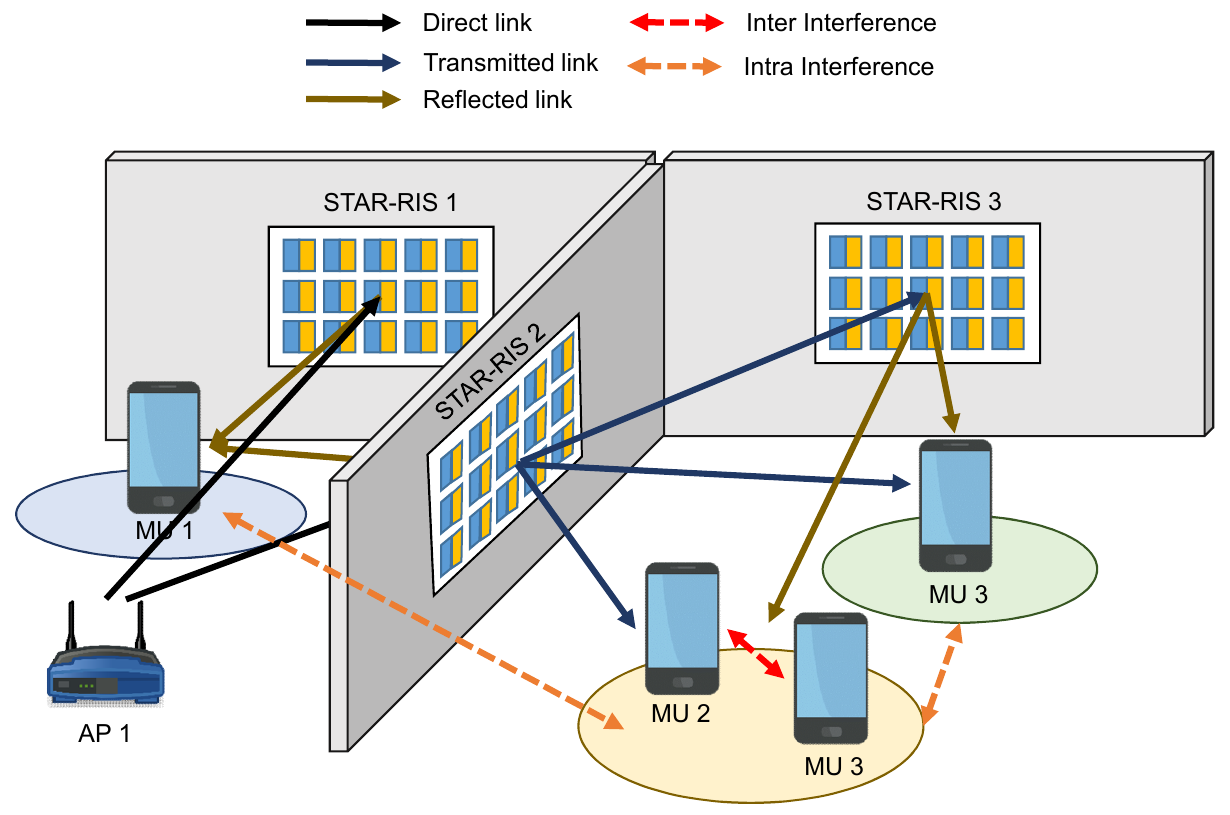}
    \caption{Illustration of our system model.}
    \label{sys_mod}
\end{figure}

As illustrated in Fig.~\ref{sys_mod}, we consider a downlink communication in Multi-STAR-RISs-aided NOMA networks for indoor environments, which consists of an access point (AP) with $N_{b}$ antenna, a set $\mathcal{L}$ of $L$ STAR-RISs, and a set $\mathcal{U}$ of $U$ mobile users (MUs) with a single antenna. We also consider STAR-RISs with $M=M_{h}M_{v}$ elements, where $M_{h}$ and $M_{v}$ denote the number of elements along the vertical and the horizontal, respectively. The locations of AP, center of STAR-RIS $l \in \mathcal{L}$ and MU $u \in \mathcal{U}$ are ${p}_{0}=\left[{x}_{0},{y}_{0},h_{0}\right]^{T}$, ${p}_{l}=\left[{x}_{l},{y}_{l},h_{l}\right]^{T}$ and ${p}_{u}=\left[{x}_{u},{y}_{u}\right]^{T}$. As shown in Fig.~\ref{sys_mod}, we assume that the environment under consideration consists of several rooms, and some of the walls of the rooms are composed of START-RIS. Downlink communication between APs and MUs in different rooms is possible through walls composed of STAR-RISs. In this paper, we assume that perfect channel state information (CSI) is available at the AP to investigate the performance gain of Multi-STAR-RISs. The STAR-RIS adopts an energy splitting (ES) protocol, where each element can operate simultaneous transmission and reflection modes. For given transmission and reflection amplitude coefficients, the signals incident upon each element is split into transmitted and reflected signals having different energy. In a practical implementation, the amplitude and phase shift coefficients of each element for transmission and reflection will be jointly optimized for achieving diverse design objectives in wireless networks.

In our system model, the reflection and transmission surfaces of STAR-RIS are considered differently depending on the location of the AP. Thus, we assume that clockwise surfaces are forward and otherwise backward. Therefore, the forward and backward side passive beamforming vectors of STAR-RIS $l$ are given by
\begin{equation}
\boldsymbol{\Phi}^{F}_{l} = \textrm{diag}\left\{\sqrt{\beta^{F}_{l_1}} e^{j\theta^{F}_{l_1}}, \sqrt{\beta^{F}_{l_2}} e^{j\theta^{F}_{l_2}}, ..., \sqrt{\beta^{F}_{l_M}} e^{j\theta^{F}_{l_M}} \right\},\label{eq_Phi_t_l}
\end{equation}
\begin{equation}
\boldsymbol{\Phi}^{B}_{l} = \textrm{diag}\left\{\sqrt{\beta^{B}_{l_1}} e^{j\theta^{B}_{l_1}}, \sqrt{\beta^{B}_{l_2}} e^{j\theta^{B}_{l_2}}, ..., \sqrt{\beta^{B}_{l_M}} e^{j\theta^{B}_{l_M}} \right\},\label{eq_Phi_r_l}
\end{equation}
where $\beta^{F}_{l_m}$ and $\beta^{B}_{l_m}$ are amplitudes of the forward and backward sides for STAR-RIS $l$'s element $m$. $\theta^{F}_{l_m}$ and $\theta^{B}_{l_m}$ are phase shift coefficients of the forward and backward sides for STAR-RIS $l$'s element $m$.

By considering the path loss model for the indoor hotspot scenario, as presented in 3GPP TR 38.901 version 16.1.0 Release 16, the path losses, at the reference distance of the LoS link and NLoS link, are given by
\begin{equation}
PL^{\textrm{LoS}} = 32.4+17.3\log_{10}{(d)}+20\log_{10}{(f)},\label{eq_PL_LoS}
\end{equation}
\begin{equation}
PL^{\textrm{NLoS}} = 32.4+31.9\log_{10}{(d)}+20\log_{10}{(f)},\label{eq_PL_NLoS}
\end{equation}
where $d$ is the distance between the transmitter and the receiver, and $f$ is the sub-carrier frequency. The channel between an AP and a MU can be modeled as a Rician channel, which includes one LoS path and a number of NLoS paths. Therefore, the channel gain $h_{u}$ from AP to MU $u$ can be formulated as
\begin{equation}
h_{u}  = \sqrt{\frac{\kappa}{\kappa+1}}h^{\textrm{LoS}}_{u} + \sqrt{\frac{1}{\kappa+1}}h^{\textrm{NLoS}}_{u},\label{eq_h_b_u}
\end{equation}
where $\kappa$ is the Rician factor, $h^{\textrm{LoS}}_{u} \in \mathbb{C}^{N_{b} \times 1}$ and $h^{\textrm{NLoS}}_{u} \in \mathbb{C}^{N_{b} \times 1}$ are LoS and NLoS channel gains between AP and MU $u$, where $\mathbb{C}^{N_{b} \times 1}$ denotes a complex matrix of size $N_{b} \times 1$. Similarly, we can define the channel gain $g_{l}$ from AP to the STAR-RIS $l$ and the channel gain $g_{l, u}$ from STAR-RIS $l$ to MU $u$ as follows:
\begin{equation}
g_{l} = \sqrt{\frac{\kappa}{\kappa+1}}g^{\textrm{LoS}}_{l} + \sqrt{\frac{1}{\kappa+1}}g^{\textrm{NLoS}}_{l},\label{eq_g_l}
\end{equation}
\begin{equation}
g_{l,u} = \sqrt{\frac{\kappa}{\kappa+1}}g^{\textrm{LoS}}_{l,u} + \sqrt{\frac{1}{\kappa+1}}g^{\textrm{NLoS}}_{l,u},\label{eq_g_l_u}
\end{equation}
where $g^{\textrm{LoS}}_{l} \in \mathbb{C}^{N_{b} \times M}$ and $g^{\textrm{NLoS}}_{l} \in \mathbb{C}^{N_{b} \times M}$ are LoS and NLoS channel gains from AP $b$ to STAR-RIS $l$, and $g^{\textrm{LoS}}_{l,u} \in \mathbb{C}^{M \times 1}$ and $g^{\textrm{NLoS}}_{l,u} \in \mathbb{C}^{M \times 1}$ are LoS and NLoS channel gains from STAR-RIS $l$ to MU $u$. Hence, the combined channel gain from AP to MU $u$ is given by 
\begin{equation}
\hat{h}_{u} = 
\begin{cases}
h_{u} + \sum_{l\in\mathcal{L}} \left\{ c^{b}_{l}(c^{l_{F}}_{u}g_{l} \boldsymbol{\Phi}^{F}_{l} g_{l,u} + c^{l_{B}}_{u}g_{l} \boldsymbol{\Phi}^{B}_{l} g_{l,u}) \right\} \\ \hfill \text{if} \ \ {c^{b}_{u}=1}, \\
\sum_{l\in\mathcal{L}} \left\{  c^{b}_{l}(c^{l_{F}}_{u}g_{l} \boldsymbol{\Phi}^{F}_{l} g_{l,u} + c^{l_{B}}_{u}g_{l} \boldsymbol{\Phi}^{B}_{l} g_{l,u})  \right\} \\ \hfill \text{if} \ \ {c^{b}_{u}=0},
\end{cases} \label{eq_hat_h_u}
\end{equation}
where $c^{b}_{u},c^{b}_{l},c^{l_{F}}_{u},c^{l_{B}}_{u} \in \left\{ 0,1 \right\}$ are adjacency indicators between AP and MU $u$, between AP $b$ and STAR-RIS $l$, between a forward side of STAR-RIS $l_{F}$ and MU $u$, and between a backward side of STAR-RIS $l_{B}$ and MU $u$.

In NOMA networks, intra-cluster and inter-cluster interference can be considered, where intra-cluster interference occurs between MUs grouped in the same cluster of AP, and inter-cluster interference occurs between MUs grouped in different clusters of AP. Thus, the MUs associated with AP are further clustered into $K$ groups. Therefore, we define $\gamma_{k,u} \in \left\{ 0,1 \right\}$ as a user pairing factor, where $\gamma_{k,u}=1$ if MU $u$ is involved in cluster $k$ of AP, otherwise $\gamma^{b}_{k,u}=0$. Moreover, let $\omega_{b} = \left\{ w_{1}, w_{2}, ... , w_{K} \right\}$ be the active beamforming vector of AP. Therefore, the received signal of MU $u$ associated with AP in cluster $k$ can be given by
\begin{multline}
y_{k,u} = \hat{h}_{u} [\omega_{b,k}(\gamma_{k,u}p_{0}s_{k,u} + \sum^{U}_{u' \ne u} \gamma_{k,u'}p_{0}s_{k,u'}) \\ + \sum^{K}_{k' \ne k} \sum^{U}_{u'} \gamma_{k',u'}\omega_{k'}p_{0}s_{k',u'}] + N_{0},\label{eq_y_b_u}
\end{multline}
where \textcolor{black}{$p_{0}$ is the power allocation coefficient of each MU associated with AP. We assume that MUs connected to the AP use power equally. Therefore, the power allocation coefficient for AP satisfies $p_{0}=1/|\mathcal{U}|$.} $s_{k,u}$ denotes the signal transmitted by AP for MU $u$ in cluster $k$, and $N_{0}$ is the Additive White Gaussian Noise (AWGN) with variance $\sigma^{2}$. Without loss of generality, for any cluster $k \in \mathcal{K}$, $\delta(u)$ denotes the MU index that corresponds to MU $u$ decoded order in the Successive Interference Cancellation (SIC) procedure. For cluster $k$, after applying the SIC decoding procedure \cite{cui2018unsupervised}, the intra-cluster and inter-cluster powers of MU $u$ associated with AP on cluster $k$ can be calculated as
\color{black}
\begin{equation}
I^{\textrm{intra}}_{k,u} =|\hat{h}_{u}\omega_{k}|^{2}\sum^{U}_{\delta_{k}(u') > \delta_{k}(u)} \gamma_{k,u'}p_{0}, \label{eq_intra}
\end{equation}
\begin{equation}
I^{\textrm{inter}}_{k,u} = \sum^{K}_{k' \ne k} \sum^{U}_{u'}\gamma_{k',u'}|\hat{h}_{u}\omega_{k'}|^{2} , \label{eq_inter}
\end{equation}

Accordingly, the received signal-to-interference-plus-noise ratio (SINR) of MU $u$ associated with AP in cluster $k$ is given by
\begin{equation}
\textrm{SINR}_{k,u} = {\frac{|\hat{h}_{u}\omega_{k}|^{2}\gamma_{k,u}p_{0}}{I^{\textrm{intra}}_{k,u} + I^{\textrm{inter}}_{k,u} + \sigma^2}}, \label{eq_sinr}
\end{equation}

For any two MUs $v$ and $u$ with decoding order $\delta_{k}(v) > \delta_{k}(u)$ in the same AP and cluster $k$, the received SINR of the signal $s_{k,u}$ at the MU $v$ is given by
\begin{equation}
\textrm{SINR}_{k,v \rightarrow u} = {\frac{|\hat{h}_{b,v}\omega_{k}|^{2}\gamma_{k,v}p_{0}}{I^{\textrm{intra}}_{k,v \rightarrow u} + I^{\textrm{inter}}_{k,v \rightarrow u} + \sigma^2}}, \label{eq_sinr_v_u}
\end{equation}
where $I^{\textrm{intra}}_{k,v \rightarrow u}=|\hat{h}_{v}\omega_{k}|^{2}\sum^{U}_{\delta_{k}(u') > \delta_{k}(u)} \gamma_{k,u'}p_{0}$ is the intra-cluster interference power of the signal $s_{k,u}$ at MU $v$. $I^{\textrm{inter}}_{k,v \rightarrow u} = \sum^{K}_{k' \ne k} \sum^{U}_{u'} \gamma_{k',u'}|\hat{h}_{v}\omega_{k'}|^{2}$ is the inter-cluster interference power of the signal $s_{k,u}$ at MU $v$. It is worth pointing out that given a decoding order, to guarantee the SIC performed successfully, the condition $\textrm{SINR}_{k,v \rightarrow u} \geq \textrm{SINR}_{k,u}$ with $\delta_{k}(v) > \delta_{k}(u)$ must be guaranteed. Therefore, the achievable data rate of MU $u$ associated with AP in cluster $k$ is calculated as
\begin{equation}
R_{k,u} = \log_2 \left( 1 + \textrm{SINR}_{k,u} \right), \label{eq_R_u}
\end{equation}

\subsection{Problem Formulation}
\label{prob_form}
In this subsection, we define the detailed problem formulation based on the proposed system model. This work's major goal is to maximize the achievable sum rate of $U$ MUs (considered as a network utility), while jointly optimizing user pairing factor $\boldsymbol{\gamma}$, decoding order $\boldsymbol{\delta}$, active beamforming $\boldsymbol{\omega}$, and passive beamforming $\boldsymbol{\Phi}=\left\{\Phi^{F}, \Phi^{B}\right\}$ of STAR-Multi-RISs. Therefore, we can define our optimization problem as follows:
\begin{maxi!}[2] 
    {\substack{\boldsymbol{\gamma}, \boldsymbol{\delta}, \boldsymbol{\omega}, \boldsymbol{\Phi}}}   
    {\sum_{k=1}^{K} \sum_{u=1}^{U}  R_{k,u}}{\label{opt:P1}}{\textbf{P1:}}
    \addConstraint{R_{k,u} \geq R^{\textrm{min}}_u,~\forall u \in \mathcal{U} \label{P1_C1}}
    \addConstraint{
    \textrm{SINR}_{k,v \rightarrow u} \geq \textrm{SINR}_{k,u},~\delta_{k}(v) > \delta_{k}(u)
    \label{P1_C9}}
    \addConstraint{
    \gamma_{k,u} \in \left\{0,1\right\},~\forall k \in \mathcal{K},~\forall u \in \mathcal{U}
    {\label{P1_C2}}}
    \addConstraint{ \sum_{k=1}^{K} \lVert w_{k} \rVert^{2} \leq P_{\textrm{max}} \label{P1_C5}}
    \addConstraint{ \sqrt{\beta^{t}_{l_m}}, \sqrt{\beta^{e}_{l_m}} \in [0,1],~\forall l \in \mathcal{L},~\forall m \in \mathcal{M} \label{P1_C6}}
    \addConstraint{ \beta^{t}_{l_m} + \beta^{r}_{l_m} = 1,~\forall l \in \mathcal{L},~\forall m \in \mathcal{M} \label{P1_C7}}
    \addConstraint{ \theta^{t}_{l_m},\theta^{r}_{l_m}\in [0,2\pi),~\forall l \in \mathcal{L},~\forall m \in \mathcal{M}, \label{P1_C8}}
\end{maxi!}
\textcolor{black}{where $R^{\textrm{min}}_u$ is the minimum rate requirement of each MU. Constraint (\ref{P1_C1}) guarantees the QoS requirement of each MU, and constraint (\ref{P1_C9}) ensures the success of the SIC decoding. Furthermore, constraint (\ref{P1_C2}) represents the binary variables. Constraint (\ref{P1_C5}) ensures the power budget constraint of each AP. Finally, Constraints (\ref{P1_C6}) to (\ref{P1_C8}) indicate the requirements of each reflecting and transmission element in STAR-RIS. To solve this proposed problem, we provide a solution approach in the next section.}

\section{Solution Approach}
\label{sol_algo}
As our proposed problem (\ref{opt:P1}) is a mixed-integer non-linear programming (MINLP), which is NP-hard due to complexity. To solve this problem, we first introduce the decoding order for NOMA networks. Next, we decompose the main problem into two sub-problems by using the block coordinate descent (BCD) method, and then solve each sub-problem iteratively until the convergence criteria meet.

\subsection{Decoding Order}
\label{sol_decoding}
Prior to handling the pairing and beamforming optimization problems, the decoding order must be addressed because it is a important one for the Multi-STAR-RIS in NOMA networks. Therefore, we propose a scheme to obtain the optimal decoding order by the following lemma.\\
\textbf{Lemma 1.}~\textit{Given the active beamforming vector $\boldsymbol{\omega}$ and the passive beamforming vector $\boldsymbol{\Phi}$, The decoding order for cluster $k$ with $|C_{k}|$ MUs in AP is defined as
\begin{equation}
g^{k}_{\delta_{k(1)}} \leq g^{k}_{\delta_{k(2)}} \leq \cdots \leq g^{k}_{\delta_{k}(|C_{k}|)}, \label{eq_decoding_order}
\end{equation}
where $g^{k}_{\delta_{k(j)}}={\frac{|\hat{h}_{u}\omega_{k}|^{2}}{\sum^{K}_{k' \ne k} \sum_{u'\in C_{k'}}|\hat{h}_{u}\omega_{k'}|^{2} + \sigma^2}}$  is the equivalent-combined channel gain \cite{zuo2022joint, higuchi2013non}.}

\textbf{Lemma 1} indicates that the decoding order for each cluster of the Multi-STAR-RIS in the NOMA system is a function of the active beamforming vectors $\boldsymbol{\omega}$, the passive beamforming vectors $\boldsymbol{\Phi}$.\\
\textbf{Proposition 1.}~\textit{For any two users $u$ and $v$ belong to cluster $k$, if the decoding order of the two users satisfies
\begin{equation}
\delta_{k}^{-1}(v) > \delta_{k}^{-1}(u),
\end{equation}
where $\delta_{k}^{-1}(\cdot)$ is the inverse of mapping function $\delta_{k}(\cdot)$. Then, under the optimal decoding order, the following SIC condition is guaranteed:}
\begin{equation}
\textrm{SINR}_{k,v \rightarrow u} \geq \textrm{SINR}_{k,u}.
\end{equation}
According to \textbf{Proposition 1}, the constraint in (\ref{P1_C9}) can be removed under the optimal decoding order of the NOMA system. This operation will not affect the optimality of the problem (\ref{opt:P1}). Furthermore, \textbf{Lemma 1} and \textbf{Proposition 2} guarantee that once the association, pairing, and beamforming vectors are determined, the optimal decoding order in each cluster is fixed \cite{zuo2022joint, cui2018unsupervised}. Therefore, we develop the optimal beamforming vectors for MUs in STAR-Multi-RIS NOMA system based on this observation.

\subsection{Correlation-Based K-means Clustering for MU Pairing} 
\label{sol_clustering}
\begin{algorithm}[t]
	\caption{\strut Correlation-based K-means Clustering for MU Pairing} 
	\label{alg:pairing}
	\begin{algorithmic}[1]
	    \STATE{\textbf{Input:} the initial passive beamforming vector $\boldsymbol{\Phi}_0$}
        \STATE{$C_{\textrm{pre}}, C = \emptyset$}
        \STATE{Randomly select $\chi_{k}$ from $u \in \mathcal{U}$ , $\forall k=1,...,K$}
        \WHILE{$C_{\textrm{pre}} \neq C$}
        \STATE{$C_{\textrm{pre}} \leftarrow C$}
        \FOR{$u \in \mathcal{U}$ with $u \neq \chi_{k}$ , $\forall k=1,...,K$}
        \STATE{$k^{*}=\arg\max_{1 \leq k^{*} \leq K}C_{u,\chi_{k^{*}}}$}
        \STATE{$C_{k^{*}}=C_{k^{*}} \cup \left\{ u \right\}$}
        \ENDFOR
        \STATE{Update $\chi_{k}$ according to (\ref{eq_representative}), $\forall k=1,...,K$}
        \STATE{$C_{k} = C_{k} \setminus u$, $\forall u \neq \chi_{k} \in C_{k}$, $\forall k=1,...,K$}
        \ENDWHILE
		\STATE{\textbf{Output:} The optimal clustering vector $\mathbf{C}^{*} \rightarrow$ MU pairing vector $\boldsymbol{\gamma}^*$.}
	\end{algorithmic}
\end{algorithm}

MUs whose channels are highly correlated should be assigned to the same group to make full use of the multiplexing gain, while MUs whose channels are uncorrelated should be assigned to different groups to decrease the interference. We adapt the K-means clustering algorithm to implement the MU pairing in NOMA networks. K-means clustering is one way to divide given data into multiple partitions \cite{hartigan1979algorithm}. The K-means algorithm determines the cost function as the sum of squares at the center of each group and the group's distance from the data subject. Clustering is also performed by updating the group that each data object belongs to to minimize the value of this cost function. Therefore, we use the channel correlation between each MU for the cost function for MU pairing. The normalized channel correlation between MU $i$ and MU $j$ can be calculated as \cite{zhu2019millimeter}

\begin{equation}
\textrm{Cor}_{i,j} = {\frac{\hat{h}^{\textrm{H}}_{i}\hat{h}_{j}}{\lVert \hat{h}_{i} \rVert \lVert \hat{h}_{j} \rVert}}. \label{eq_correlation}
\end{equation}

In order to cluster MUs connected to each AP $b$ into $K_b$, we select randomly $K_b$ MUs assigned to clusters, $C = \left\{C_{1}, C_{2},..., C_{K}\right\}$, one by one. Then, the channel correlation between the unselected MU $u'$ and the selected MU $u$ is calculated based on (\ref{eq_correlation}), and the MU $u'$ having the highest channel correlation is assigned to the cluster to which the MU $u$ belongs. Thenceforth, each representative can be selected from each cluster. The representative of each cluster is updated as the one with the lowest correlation with the other clusters in order to further reduce the correlation of the channels between the various clusters. The correlation between a MU to the other clusters is the total normalized channel correlation between a MU to the MUs of the other clusters. The correlation between a MU $u$ in the cluster $C_k$ to the other clusters is defined as

\begin{equation}
\Bar{\textrm{Cor}}_{u} = \sum^{K}_{l \neq k} \sum_{u' \in C_{l}} \textrm{Cor}_{u,u'}. \label{eq_correlation_cluster}
\end{equation}

After that, the representative $\chi_{k}$ of the cluster $C_k$ is updated as

\begin{equation}
\chi_{k} = \arg\min_{u \in C_{k}} \Bar{\textrm{Cor}}_{u}. \label{eq_representative}
\end{equation}

Following the update of the representative for each cluster, the other MUs are subsequently reassigned to their respective clusters. The iteration is terminated when the representatives of the clusters remain unaltered. Finally, the optimal output $\mathbf{C}^{*}$ of the clustering vector, is transformed to the MU pairing vector $\boldsymbol{\gamma}^{*}$. The Correlation-based K-means Clustering is described in Algorithm \ref{alg:pairing}.

\begin{algorithm}[t]
	\caption{\strut Learning Process for Multi-Agent Proximal Policy Optimization (MAPPO)} 
	\label{alg:MAPPO}
	\begin{algorithmic}[1]
	    \STATE{\textbf{Initialize:} the initial network $\pi^{\omega}_{0}$ and $\pi^{\Phi}_{0}$ for agent of active and beamforming.}
		\FOR{episode$=1,2,...,E$}
		\STATE{Initialize randomly each MU's position and calculate the optimal user pairing from \ref{sol_clustering} solution, i.e., $\boldsymbol{\gamma}^{*}$.}
		\FOR{time slot$=1,2,...,N$}
		\STATE{Update observation $\mathcal{S}(n)$.}
		\STATE{Run policy $\mathcal{A}^{\omega}\sim\pi^{\omega}_{\theta_{\textrm{old}}}, \mathcal{A}^{\Phi}\sim\pi^{\Phi}_{\theta_{\textrm{old}}}$.}
		\STATE{Compute the common reward $\mathcal{R}(n)$.}
		\STATE{Save $(\mathcal{S}(n),\mathcal{A}(n),\mathcal{R}(n),\mathcal{S}(n+1))$ in memory of each agent.}
		\ENDFOR
		\STATE{Compute advantage estimates $\left\langle \hat{A}^{\omega}_{1},...,\hat{A}^{\omega}_{N} \right\rangle$, $\left\langle \hat{A}^{\Phi}_{1},...,\hat{A}^{\Phi}_{N} \right\rangle$  based on (\ref{eq_adv}).}
		\STATE{Optimize surrogate $L^{\textrm{PPO}}$ wrt $\theta^{\omega}$, $\theta^{\Phi}$ with minibatch from memory based on (\ref{L_PPO}).}
		\STATE{$\theta^{\omega}_{\textrm{old}} \leftarrow \theta^{\omega}$, $\theta^{\Phi}_{\textrm{old}} \leftarrow \theta^{\Phi}$}
		\ENDFOR
		\STATE{\textbf{Output:} Optimal networks $\pi^{\omega}_{\theta_{\textrm{opt}}}$, $\pi^{\Phi}_{\theta_{\textrm{opt}}}$.}
	\end{algorithmic}
\end{algorithm}

\subsection{Multi-Agent Reinforcement Learning (MARL) Scheme}
 At the fixed MU association, pairing, and decoding order, we can rewrite the beamforming vectors optimization problem as follows:
\label{sol_marl}
\begin{maxi!}[2] 
		{\substack{\boldsymbol{\omega}, \boldsymbol{\Phi}}}   
		{\sum_{k=1}^{K} \sum_{u=1}^{U} R_{k,u}}{\label{opt:P1.1}}{\textbf{P1.1:}}
		\addConstraint{\textcolor{black}{({\ref{P1_C1}}),({\ref{P1_C5}})\sim({\ref{P1_C8}}).}}
\end{maxi!}
However, the subproblem $\bold{P1.1}$ has the non-convex nature. Thus, it is challenging to solve using the existing optimization techniques. Thus, we propose a proximal policy optimization (PPO)-based multi-agent reinforcement learning method (MARL) to solve problem $\bold{P1.1}$. PPO is a widely used reinforcement learning technique that is known for its simplicity in implementation and applicability across diverse situations \cite{ppo}. Furthermore, it has demonstrated consistent and reliable performance. The utilization of the Proximal Policy Optimization (PPO) technique serves to streamline the intricate computational process associated with trust region policy optimization (TRPO). The Trust Region Policy Optimization (TRPO) algorithm aims to optimize a surrogate objective function in the following manner \cite{trpo}.

\begin{figure*}[t!]
        \begin{subfigure}[t]{0.33\textwidth}
               \captionsetup{justification=raggedright,singlelinecheck=false}
                \includegraphics[width=\linewidth]{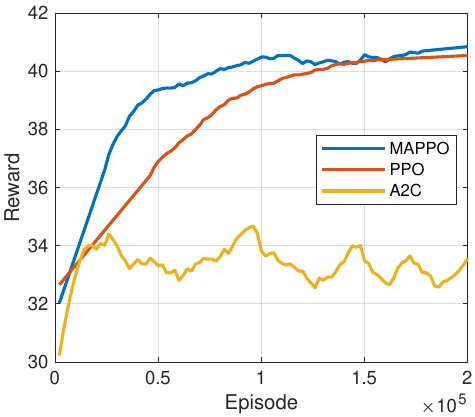}
                \caption{Learning convergence results based on different algorithms.}
                \label{simulation_result_1}
        \end{subfigure}%
        \begin{subfigure}[t]{0.33\textwidth}
               \captionsetup{justification=raggedright,singlelinecheck=false}
               \includegraphics[width=\linewidth]{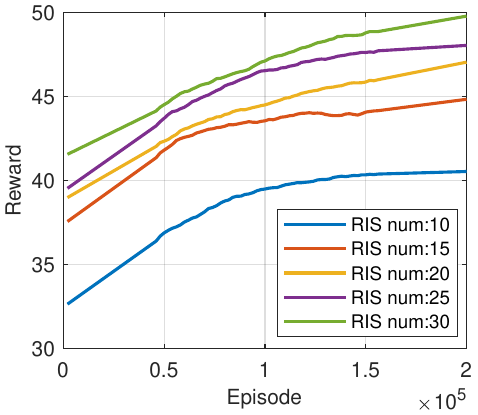}
                \caption{Learning results based on the number of STAR-RIS elements.}
                \label{simulation_result_2}
        \end{subfigure}%
        \begin{subfigure}[t]{0.33\textwidth}
         \captionsetup{justification=raggedright,singlelinecheck=false}
         \includegraphics[width=\linewidth]{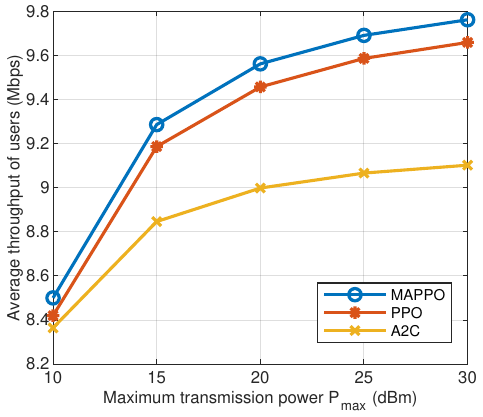}
                \caption{Average throughput of MUs with maximum transmission power changes.}
                \label{simulation_result_3}
        \end{subfigure}%
        \caption{Simulation results of proposed scheme convergence and comparison with baselines.}\label{mixed1}
\end{figure*}

\begin{figure}[t]
    \centering
    \includegraphics[width=7.5cm]{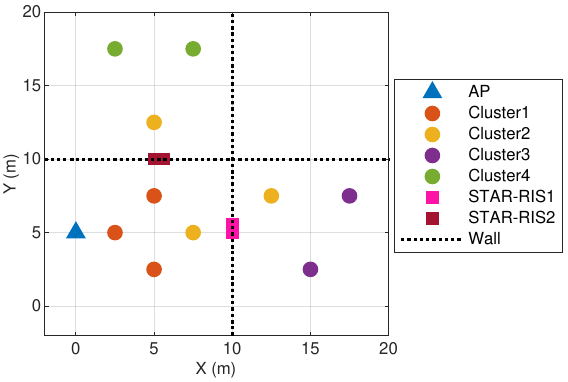}
    \caption{Deployments of AP, STAR-RISs, MUs.}
    \label{fig_deply}
\end{figure}

\begin{equation}
L^{\textrm{TRPO}}_{n}(\theta)=\hat{\mathbb{E}}_{n} \left[ {\frac{\pi_{\theta}(\mathcal{A}_{n}|\mathcal{S}_{n})}{\pi_{\theta_{\textrm{old}}}(\mathcal{A}_{n}|\mathcal{S}_{n})}} \hat{A}_{n}  \right] = \hat{\mathbb{E}}_{n} \left[ r_{n}(\theta)\hat{A}_{n} \right],\label{L_TRPO}
\end{equation}
where $r_{n}(\theta)$ denotes the probability ratio, $\mathcal{A}_{n}$ and $\mathcal{S}_{n}$ are an action and reward in time step $n$. The surrogate objective function $L^{\textrm{TRPO}}$ of TRPO has a complicated formula expansion and must calculate the second derivative. Therefore, maximization of the surrogate objective function $L^{\textrm{TRPO}}$ of TRPO would result in an unnecessarily large policy update in the absence of a constraint. Hence, in PPO, the limitations of TRPO were addressed by incorporating an approximation of the first derivative using the clipping technique. The following is the objective function to which the clipping is applied:
\begin{equation}
L^{\textrm{CLIP}}_{n}(\theta)=\hat{\mathbb{E}}_{n} \left[ \textrm{min}(r_{n}(\theta)\hat{A}_{n}, \textrm{clip}(r_{n}(\theta), 1-\epsilon, 1+\epsilon)\hat{A}_{n}) \right],\label{L_CLIP}
\end{equation}
where $\epsilon$ is a hyperparameter and $\hat{A}_{n}$ is a truncated version of generalized advantage estimation which can be defined as follows:
\begin{equation}
    \hat{A}_{n} = \delta_{n} + (\gamma\lambda)\delta_{n+1} + \dots + (\gamma\lambda)^{N-n+1}\delta_{N-1},
\label{eq_adv}
\end{equation}
where $\delta_{n}=r_{n}+\gamma V(s_{n+1}) - V(s_{n})$. The function in \eqref{L_CLIP} takes a lower value when comparing the objectives used in the TRPO with the objectives to which clipping is applied. With this clipping method, we only consider the change in the probability ratio if it improves the objective. If it makes the objective worse, we leave it out.

\begin{figure*}[t!]
        \begin{subfigure}[t]{0.32\textwidth}
               \captionsetup{justification=raggedright,singlelinecheck=false}
                \includegraphics[width=\linewidth]{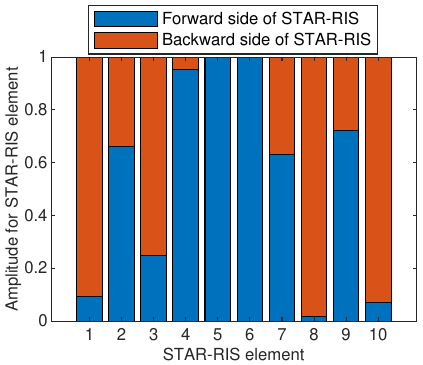}
                \caption{Amplitude by Element for STAR-RIS 1 in Fig.~\ref{fig_deply}.}
                \label{simulation_result_4}
        \end{subfigure}%
        \begin{subfigure}[t]{0.32\textwidth}
               \captionsetup{justification=raggedright,singlelinecheck=false}
               \includegraphics[width=\linewidth]{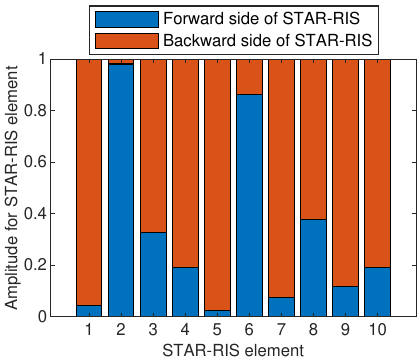}
                \caption{Amplitude by Element for STAR-RIS 2 in Fig.~\ref{fig_deply}.}
                \label{simulation_result_5}
        \end{subfigure}%
        \begin{subfigure}[t]{0.32\textwidth}
         \captionsetup{justification=raggedright,singlelinecheck=false}
         \includegraphics[width=\linewidth]{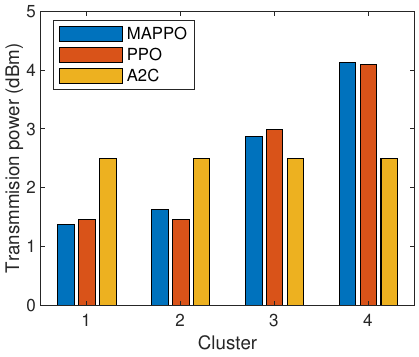}
                \caption{Status of active beamforming allocation for clusters by algorithm.}
                \label{simulation_result_6}
        \end{subfigure}%
        \caption{Analysis of optimal values derived from the learning model.}\label{mixed2}
\end{figure*}

Subsequently, PPO incorporates an actor-critic network architecture, wherein the policy and value functions share the parameters inside the network design. In the context of utilizing a neural network architecture to share parameters between the policy function and the value function, it is imperative to deploy a loss function that effectively integrates the policy surrogate and an error term derived from the value function. The objective is achieved by integrating the policy surrogate with a value function error component as follows:
\begin{equation}
L^{\textrm{PPO}}_{n}(\theta)=\hat{\mathbb{E}}_{n} \left[ L^{\textrm{CLIP}}_{n}(\theta) - c_{1} L^{\textrm{VF}}_{n}(\theta) + c_{2}E[\pi_\theta](s_n) \right],\label{L_PPO}
\end{equation}
where $c_1$ and $c_2$ are coefficients, $E$ denotes an entropy function, and $L^{\textrm{VF}}_{n}$ is a squared-error loss. In this objective $L^{\textrm{PPO}}$ can further be augmented by adding an entropy bonus $E$ to ensure sufficient exploration. 

The proposed MAPPO in this paper is each optimization variable ($\boldsymbol{\omega}, \boldsymbol{\Phi}$) PPO agents. Each agent learns simultaneously in the same environment. Moreover, each agent can determine the optimal action for a common reward. Therefore, we introduce the Markov Decision Process (MDP) of the agents used for learning.
\begin{equation}
\mathcal{S}(n) =  \{ \boldsymbol{\gamma}^{*}, \boldsymbol{\omega}(n-1), \boldsymbol{\Phi}(n-1), \{ \hat{h}_{u} \}_{u \in \mathcal{U}} \} \label{eq_state},
\end{equation}
\begin{equation}
\mathcal{A}^{\omega}(n)=\{\omega_{k}(n)\}_{k\in \mathcal{K}},
\end{equation}
\begin{equation}
\mathcal{A}^{\Phi}(n)=\{\Phi^{F}_{l}, \Phi^{B}_{l}\}_{l\in \mathcal{L}},
\end{equation}
\begin{equation}
\mathcal{R}(n)= \textrm{min}_{u \in \mathcal{U}}(R_{k,u}),
\end{equation}
where $\boldsymbol{\omega}(n-1)$ and $\boldsymbol{\Phi}(n-1)$ denote the active and passive beamforming vector at the last step, $\mathcal{A}^{\omega}(n)$ and $\mathcal{A}^{\Phi}(n)$ denote the active and passive beamforming agent. The reward is determined by the lowest throughput of all users. This is to satisfy the minimum throughput of all users for the constraint (\ref{P1_C1}). Based on the MDP for the proposed optimization problem, the optimal value of the decision variable can be obtained by executing the proposed MAPPO algorithm, such as Algorithm \ref{alg:MAPPO}.

\section{Simulation Results}
\label{simul}
We consider one AP in a region of $20~$m$^2$, which provides services to the $10$ MUs in our simulation setup. Then, the considered region is divided into four rooms, each of which is surrounded by walls. Furthermore, two STAR-RISs are installed in the room where the AP is located. In the learning stage, MUs are randomly placed, and in the verification stage, MUs are placed as in Fig.~\ref{fig_deply}. The AP has four antennas with a frequency of $6$ GHz and a noise density of $-100$ dBm/Hz. Each STAR-RIS is composed of 10 elements, and separations between elements are $0.2$ in width and $0.1$ in length. One episode has $10$ time slots for learning settings, and the learning network model has $2$ hidden layers and $256$ units for each hidden layer. In our simulation, correlation-based K-means clustering for MU pairing is performed after MUs are assigned at the beginning of each episode. Finally, to evaluate the performance of the proposed algorithm, we use the following three different algorithms as follows:
\begin{itemize}
    \item \emph{MAPPO (proposed)}: \textcolor{black}{The proposed MAPPO algorithm, in which each beamforming vector variable is thought of as an agent that learns its variable optimization.}
    \item \emph{PPO}: The general PPO algorithm learns all beamforming vector variables in one network at a time.
    \item \emph{A2C}: The multi-agent advanced actor-critic (MAA2C) algorithm, in which each beamforming vector variable is thought of as an agent that learns its variable optimization.
\end{itemize}

Fig.~\ref{simulation_result_1} shows the results of learning convergence based on different algorithms. Our proposed MAPPO completed the learning with the fastest and highest rewards, followed by a slow PPO but similar rewards. A2C showed learning results that failed to converge and continued to vibrate.
Fig.~\ref{simulation_result_2} shows the results of learning convergence according to the number of elements in STAR-RIS. It can be seen that the final reward increases as the number of STAR-RIS elements increases, but at the same time, it can be seen that the improvement of the reward becomes smaller and smaller. This shows that the efficiency of learning can vary depending on the requirements of STAR-RIS because learning was done with an artificial neural network of the same size.
Fig.~\ref{simulation_result_3} compares the average throughput of users by algorithm according to the change in maximum transmission power. As checked in Fig.~\ref{simulation_result_1}, it was confirmed that the proposed MAPPO showed the best performance according to the results of learning convergence, followed by PPO and A2C.

Also, we record and present optimal values in a fixed environment for an in-depth understanding of optimization. Fig.~\ref{simulation_result_4} and Fig.~\ref{simulation_result_5} show the amplitude of STAR-RIS by element in the environment of Fig.~\ref{fig_deply}. It is noteworthy that to optimize the communication throughput of MUs, STAR-RIS assigns large amplitudes where there are no APs. Adding the amplitudes of all elements, STAR-RIS 1 was assigned more amplitudes in the opposite direction of the AP with $5.3997595:4.6002405$, and likewise, STAR-RIS 2 $3.19707643:6.802357$ was assigned more amplitudes in the opposite direction of the AP. It can be confirmed that this is an optimal choice of STAR-RIS for improving overall communication performance.
Finally, Fig.~\ref{simulation_result_6} shows how much active beamforming was allocated to the cluster by algorithms in the environment of Fig.~\ref{fig_deply}. The proposed MAPPO allocated relatively little power to Cluster $1$ and Cluster $2$ because users close to AP can be guaranteed sufficient communication throughput over distance. On the contrary, Cluster $3$ and Cluster $4$ are users who do not have a direct connection to the AP, so they are optimized to ensure minimum throughput by allocating high power. The PPO algorithm showed similar allocations to the MAPPO, but the A2C algorithm allocates similar power for all clusters, which can be inferred from the results of Fig.~\ref{simulation_result_3} above that the performance was low.

\section{Conclusions}
In this paper, we have studied the joint design and optimization of Multi-STAR-RISs-aided NOMA in an indoor environment using MARL. Then, we formulated an optimization problem to maximize the total throughput of MUs by optimizing user pairing, active beamforming vector, and passive beamforming vector while satisfying the resource constraints. We have divided the original problem into two subproblems to address this problem. Firstly, we have employed correlation-based K-means clustering to solve the MU pairing problem. Then, to jointly deal with beamforming vector optimizations, we have proposed the MAPPO, which can make quick decisions in the given environment owing to its low complexity. Based on the proposed MAPPO, by configuring agents for each beamforming vector, it was possible to have faster and higher performance than conventional single-agent-based PPO. In the simulation results, we have shown not only the learning convergence results for neural networks but also the learning convergence results based on the number of different STAR-RIS elements and the performance of each algorithm. In addition, while confirming how the beamforming vectors were actually optimized in one test environment, we analyzed in depth why the proposed MAPPO was high. In the future, this paper can be expanded by applying optimization through multi-cell or other metaheuristic or numerical methods that were not covered in this paper.

\label{conclusion}
\ifCLASSOPTIONcaptionsoff
  \newpage
\fi
\bibliographystyle{IEEEtran}
\bibliography{ref}
\end{document}